\documentclass[%
print,
 amsmath,amssymb,
 aps,
]{revtex4}
\usepackage[sc]{mathpazo}
\usepackage{graphicx}
\usepackage{dcolumn}
\usepackage{bm}
\usepackage{dsfont}
\usepackage[landscape,papersize={297.1mm,210mm},left=1.9cm,right=1.6cm,top=2.7cm,bottom=2.8cm]{geometry}
\usepackage[                   
            pdfstartview=FitH,
            colorlinks, 
            pdfborder=001,   
            linkcolor=blue,
            anchorcolor=blue,
            citecolor=blue,
            urlcolor=blue
            ]{hyperref}
\usepackage{amsthm}
\makeatletter

\newcommand{\Rmnum}[1]{\expandafter\@slowromancap\romannumeral #1@}
\makeatother

\begin{document}
\title{On  coherence of quantum operations by using Choi-Jamio{\l}kowski isomorphism }

\author {Xiaorong Wang$^{1}$, Ting Gao$^{1}$,}

\email{gaoting@mail.hebtu.edu.cn}
\author {Fengli Yan$^{2}$}
\email{flyan@mail.hebtu.edu.cn}
\affiliation {$^{1}$School of Mathematics Science, Hebei Normal University, Shijiazhuang 050024, China\\
$^{2}$ College of Physics, Hebei Key Laboratory of Photophysics Research and Application, Hebei Normal University,  Shijiazhuang 050024, China}

\begin{abstract}
In quantum information, most information processing processes involve  quantum channels. One manifestation of a quantum channel is quantum operation  acting on quantum states. The coherence of quantum operations can be considered as a quantum resource, which can be exploited to perform certain quantum tasks.
From the viewpoint of Choi-Jamio{\l}kowski isomorphism, we study  the coherence of quantum operations in the framework of resource theory. We define the  phase-out superoperation and give the operation which transforms the Choi-Jamio{\l}kowski state of a quantum operation  to the  Choi-Jamio{\l}kowski state of the another quantum operation  obtained by  using  the phase-out superoperation  to act on the quantum operation. The set  of maximally incoherent superoperations, the set of nonactivating coherent superoperations and the set of de-phase incoherent superoperations are defined and we prove that these sets are closed to  compound operation and convex combination of
quantum superoperations.  Further, we introduce the fidelity coherence measure of quantum operations and obtain the exact form of the fidelity coherence measure of the unitary operations on the single qubit. \\

\end{abstract}


\maketitle

\section{Introduction}

Quantum coherence is not only a fundamental feature of quantum physics, but also an important research field of quantum information theory. As a kind of special resource, it plays a significant role in quantum thermodynamics \cite{relixue1,relixue2,relixue3,relixue4}, quantum algorithm \cite{jisuanji1,jisuanji2,jisuanji3,jisuanji4}, quantum meteorology \cite{shengwu}, etc. Quantifying and applying coherence is a very interesting work.  In 2014, Baumgratz et al  established a rigorous framework of the coherence resource theory for measuring quantum coherence \cite{Baumgratz}. After that, the study on the  characterization and measure  of quantum coherence  has been made a great leap forward \cite{fazhan1,fazhan2,fazhan3,fazhan4,fazhan5, qigaoyan, fuyangao, zhanggaoyan}.

The framework of the coherence resource theory consists of three ingredients: incoherent states, incoherent operations and coherence measures. Incoherent states are those quantum states which  do not possess any coherence resource. Incoherent operations are special quantum operations which can not generate coherence from incoherent states. Coherence measures being used to quantify the coherence of quantum states are functions mapping quantum states to real numbers.

In quantum information theory, most information processing processes involve and rely on quantum channels. One of the manifestation of quantum channel is the quantum operation acting on quantum states. Therefore, some researchers  investigated the coherence of quantum operations. The works  have been done mainly on characterizing a certain property of quantum operations, such as entanglement of quantum channels \cite{jiuchan1,jiuchan2}, channel discrimination \cite{qufen1,qufen2,qufen3}, channel simulation \cite{moni1,moni2,moni3,moni5}, quantum memory \cite{jiyi1,jiyi2,jiyi3}, the coherence of quantum operation and others \cite{qita1,qita2,qita3,qita4}. As a matter of fact, the coherence of quantum operations can be regarded as a resource and applied to the corresponding quantum information processing.

Coherence resource theory of quantum operations is also composed of three elements: free operations, free superoperations and coherence measures of quantum operations. Here the superoperation is a map between operations \cite{123,52,Choi1,Choi2}. Only free superoperations can transform free operations to free operations. The coherence measures of operations are functions which map quantum operations to real numbers.

Quantifying the coherence of operations  is a very important and meaningful work which can enable us to have a deeper understanding of basic physics and also  can provide new ideas for quantum information processing. There are two main aspects in the research of the coherence of quantum operations. The first one  is to  study quantum operations from the perspective that quantum operations change the coherence of quantum states \cite{fangxiangyi1,fangxiangyi2,fangxiangyi3}.  The other one  is to  analyze quantum operations directly. Coherence resource theory of quantum operations promotes the research  strongly \cite{zhijie1,zhijie2,zhijie3,38,fenlei1,Korzekwa,Xu,Bera}. Among them,   Orzekwa et al studied the coherence of quantum operations based on entropy coherence and 2-norm coherence of quantum operations \cite{Korzekwa}.  Bera proposed a resource theory framework to quantify the superposition that exists in any quantum evolutions \cite{Bera}.  Xu defined incoherent channels and incoherent superchannels, and established a resource theory for quantifying the coherence of quantum channels \cite{Xu}.

In this paper we investigate the coherence of quantum operations from the viewpoint of Choi-Jamio{\l}kowski isomorphism. We first define the phase-out superoperation and study its properties.    Then the three sets of   superoperations  are defined, and we prove that these three sets are closed to  compound operation and convex combination of
quantum superoperations. Then we introduce the fidelity coherence measure of quantum operations, and  obtain the exact form of the fidelity coherence measure of the unitary operations on the single qubit.

\section{Preliminaries}

Before we quantify the coherence of quantum operations, an introduction of the concepts and notation that will be used in
the subsequent sections of our article is necessary.  Let $\mathcal{H}^{I}$ and $\mathcal{H}^{O}$ be two $d$-dimensional Hilbert spaces with $\{|i\rangle\}_{i=0}^{d-1}$ and  $\{|\alpha\rangle\}_{\alpha=0}^{d-1}$  being the orthonormal  bases of $\mathcal{H}^{I}$ and $\mathcal{H}^{O}$, respectively. We assume that $\{|i\rangle\}_{i=0}^{d-1}$ and $\{|\alpha\rangle\}_{\alpha=0}^{d-1}$ are fixed and adopt  the tensor basis $\{|i\alpha\rangle\}_{i\alpha}$ as the fixed basis when we consider the multipartite system  with the Hilbert space  $\mathcal{H}^{IO}=\mathcal{H}^{I}\otimes \mathcal{H}^{O}$. Let $\mathcal{O}(\mathcal{H}^{I}\rightarrow \mathcal{H}^{O})$ be the set of all  quantum operations from $\mathcal{H}^{I}$ to $\mathcal{H}^{O}$.

A quantum operation is governed  in terms of
Choi-Jamio{\l}kowski isomorphism \cite{Choi3,Choi4}. An arbitrary  operation $\Phi\in \mathcal{O}(\mathcal{H}^{I}\rightarrow \mathcal{H}^{O})$ is fully characterized by
a matrix called the Choi-Jamio{\l}kowski  matrix of the operation $\Phi$
\begin{equation}
C_{\Phi}=(\mathbb{I} \otimes \Phi)|\varphi\rangle\langle\varphi|,
\end{equation}
in $\mathcal{H}^{I}\otimes \mathcal{H}^{O}$.  Here $|\varphi\rangle=\frac{1}{\sqrt{d}} \sum_{i=0}^{d-1}|i i\rangle$ is a maximally entangled state in
$d^{2}$-dimensional  Hilbert space
$\mathcal{H}^{I} \otimes \mathcal{H}^{I}$, and $\mathbb{I}$ is the identity operation. The one-to-one correspondence  between $C_{\Phi}$ and $\Phi$ is given by
\begin{equation}
\Phi\left(\rho\right)=\operatorname{tr}_{I}\left[\left(\rho^\mathrm{T} \otimes \mathbb{I}\right) C_{\Phi}\right],
\end{equation}
where $\rho^\mathrm{T}$ is the transpose of state $\rho$. The quantum operation $\Phi$ is completely positive and trace preserving if and only if $C_{\Phi}\geq 0$ and $\mathrm{tr}_{O}(C_{\Phi})=\frac{\mathbb{I}}{d}$. When $C_{\Phi}\geq 0$ and $\mathrm{tr}_{IO}(C_{\Phi})=1$, the Choi-Jamio{\l}kowski  matrix $C_{\Phi}$ can be viewed as a density operator, a Choi-Jamio{\l}kowski quantum state in the compound space $\mathcal{H}^{I}\otimes \mathcal{H}^{O}$.  Since Choi-Jamio{\l}kowski isomorphism guarantees an equivalence between $\Phi$ and $C_{\Phi}$, so one can  treat quantum operations with the same tools which were  normally used to treat quantum states. Further, the study of the coherence of the operation in the set  $\mathcal{O}(\mathcal{H}^{I}\rightarrow \mathcal{H}^{O})$ of all quantum operations corresponding to Choi-Jamio{\l}kowski states  is transformed into the  study of  coherence of Choi-Jamio{\l}kowski state.

Assume that $\Phi\in \mathcal{O}(\mathcal{H}^{I}\rightarrow \mathcal{H}^{O})$. It has been shown that $\Phi$ is an incoherent operation (IO) if \cite{Xu}
\begin{equation}
C_{\Phi}=\sum_{i, \alpha=0}^{d-1} \frac{\Phi_{i i \alpha \alpha}}{d}|i \alpha\rangle\langle i \alpha|,
\end{equation}
where $\Phi_{i i \alpha \alpha}=d\langle i \alpha| [\mathbb{I}\otimes \Phi(|\varphi\rangle\langle\varphi|)]|i\alpha\rangle= \langle \alpha|  \Phi(|i\rangle\langle i|)|\alpha\rangle.$

Let $\mathcal{F}$ be the set of all incoherent operations in $\mathcal{O}(\mathcal{H}^{I}\rightarrow \mathcal{H}^{O})$,  $\mathcal{F}_{C}$ be the set of all Choi-Jamio{\l}kowski states of incoherent operations. A superoperation $\tilde{\Omega}$ is a map which transforms the operation $\Phi$ to the operation  $\tilde{\Omega}(\Phi)=\Phi_{1}\circ\Phi\circ\Phi_{2}$, where $\Phi_{1}, \Phi_{2}$ are also quantum operations in $\mathcal{O}(\mathcal{H}^{I}\rightarrow \mathcal{H}^{O})$. Let $\mathcal{SO}$ be the set of all superoperators i.e. $\mathcal{SO}= \{\tilde{\Omega}|\tilde{\Omega}:\mathcal{O}(\mathcal{H}^{I}\rightarrow \mathcal{H}^{O})\rightarrow \mathcal{O}(\mathcal{H}^{I}\rightarrow \mathcal{H}^{O})\}$.  A superoperation can be understood as a quantum operation that relates input Choi-Jamio{\l}kowski states and output Choi-Jamio{\l}kowski states, with an associated operator-sum representation. That is, for a superoperation $\tilde{\Omega}$ relating input $\Phi$ and output $\Lambda$ operations, one may write
\begin{equation}
C_{\Lambda}=\Omega\left(C_{\Phi}\right)
=\sum_{n} K_{n} C_{\Phi}K_{n}^{\dagger}
=\sum_{n} p_{n} C_{\Lambda_{n}},
\end{equation}
where
\begin{equation}
p_{n}=\mathrm{tr}\left( K_{n} C_{\Phi} K_{n}^{\dagger}\right),~
C_{\Lambda_{n}}=\frac{K_{n}C_{\Phi}K_{n}^{\dagger}}{p_{n}},
\end{equation}
and $\{K_{n}\}$ are the Kraus operators of ${\Omega}$ determined by  $\tilde{\Omega}$, on the  Choi-Jamio{\l}kowski state.
Obviously, the operation $\Omega$ has a one-to-one correspondence with the superoperation $\tilde{\Omega}$.

A superoperation $\tilde{\Omega}$ is a maximally incoherent superoperation if
$\forall~ \Phi \in \mathcal{F}$ one has $\tilde{\Omega}(\Phi)\in \mathcal{F}$.  We call  $\tilde{\Omega}$   an
incoherent superoperation if Kraus operators $\{K_{m}\}$ of $\Omega$ determined by  $\tilde{\Omega}$ are  incoherent for each $m$ \cite{Xu}. We use  $\mathcal{MISO}$ and  $\mathcal{ISO}$ to denote the set of all maximally incoherent superoperations and  the set of all incoherent superoperations, respectively.

A coherence  measure $M$ of quantum operations should satisfy  the following  conditions \cite{Xu}.

(1) Nonnegativity: $M(\Phi)\geq 0$, for any $\Phi\in \mathcal{O}(\mathcal{H}^{I}\rightarrow \mathcal{H}^{O})$;~and $M(\Phi) = 0$ if and only if $\Phi\in \mathcal{F}$.

(2) Monotonicity: $M\left(\tilde{\Omega}(\Phi)\right) \leq M(\Phi)$, for any $\tilde{\Omega} \in \mathcal{ISO}$, $\Phi\in \mathcal{O}(\mathcal{H}^{I}\rightarrow \mathcal{H}^{O})$.

(3) Strong monotonicity: $\sum_{n} p_{n} M\left(\Lambda_{n}\right)\leq M(\Phi)$ for any $\tilde{\Omega}\in \mathcal{ISO}$, with $\{K_{n}\}$  being an incoherent expression of $\Omega$ corresponding to $\tilde{\Omega}$, $p_{n}=\operatorname{tr}\left(K_{n} C_{\Phi} K_{n}^{\dagger}\right)$,~and $C_{\Lambda_{n}}=\frac{K_{n} C_{\Phi}K_{n}^{\dagger}}{p_{n}}$.

(4) Convexity: $M\left(\sum_{n} p_{n} \Phi_{n}\right) \leq \sum_{n} p_{n} M\left(\Phi_{n}\right)$, where $\{\Phi_{n}\}$ are quantum operations belonging to $\mathcal{O}(\mathcal{H}_{I}\rightarrow \mathcal{H}_{O})$, and $\{p_{n}\}$ is a
probability distribution satisfying $p_{n}\geq 0$ and $\sum_{n}p_{n}=1$.

\section{Main results}

\subsection{The phase-out superoperation}

It is well-know that in Hilbert space there is the phase-out operation,  which  can eliminate  the coherence of the quantum state. Similarly, one can define  the phase-out superoperation \cite{Xu}. Since the incoherent states depend on the choice of the basis of the  Hilbert space, so does the phase-out operation.

{\bf{Definition  1}}.  A superoperation $\tilde{\Theta}\in \mathcal{SO}$ is called the phase-out superoperation  if  $\tilde{\Theta}(\Phi)$ is a incoherent operation for any quantum operation $\Phi$. The phase-out superoperation  is implemented by
\begin{equation}\begin{aligned}
\tilde{\Theta}(\Phi)
=\Delta^{O}\circ\Phi\circ\Delta^{I},
\end{aligned}\end{equation}
where $\Delta^{O}$ and $\Delta^{I}$~are
the phase-out operations on Hilbert space $\mathcal{H}^{O}$ and $\mathcal{H}^{I}$ respectively.

According to  Definition $1$, we  have

{\bf{Theorem} 1.1}. The specific  form of the quantum operation $\Theta$ corresponding to the phase-out superoperation $\tilde{\Theta}$ is
\begin{equation}\begin{aligned}
C_{\tilde{\Theta}(\Phi)}=\Theta(C_{\Phi})=\sum_{i, \alpha=0}^{d-1}\langle i \alpha|C_{\Phi}| i \alpha\rangle|i \alpha\rangle\langle i \alpha|.
\end{aligned}\end{equation}

{\bf{Proof}}. For any quantum operation $\Phi\in \mathcal{O}(\mathcal{H}^{I}\rightarrow \mathcal{H}^{O})$, the Choi-Jamio{\l}kowski state corresponding to $\Phi$ is
\begin{equation}
C_{\Phi}=\sum_{i,j, \alpha,\beta=0}^{d-1}\frac{\Phi_{ij\alpha \beta}}{d}|i \alpha\rangle\langle j \beta|,
\end{equation}
where $\Phi_{ij\alpha \beta}=d\langle i \alpha| [\mathbb{I}\otimes \Phi(|\varphi\rangle\langle\varphi|)]|j\beta\rangle=\langle \alpha|\Phi(|i\rangle\langle j|)|\beta\rangle$.
According to the definition of the phase-out superoperation, we know that the new operation $\tilde{\Theta}(\Phi)$ is an incoherent operation, then the Choi-Jamio{\l}kowski state corresponding to $\tilde{\Theta}(\Phi)$ must be the diagonal in the fixed basis $\{|i\alpha\rangle\}_{i,\alpha=0}^{d-1}$, i.e.,
\begin{equation}\begin{aligned}
C_{\tilde{\Theta}(\Phi)}
&=[\mathbb{I}\otimes(\Delta^{O}\circ\Phi\circ\Delta^{I})]|\varphi\rangle\langle\varphi| \\
&=\sum_{i, \alpha=0}^{d-1} \frac{\tilde{\Theta}(\Phi)_{i  i\alpha \alpha}}{d}|i \alpha\rangle\langle i \alpha|,
\end{aligned}\end{equation}
where
\begin{eqnarray}\tilde{\Theta}(\Phi)_{i  i\alpha \alpha}&= &d\langle i \alpha| \mathbb{I}\otimes \tilde{\Theta}(\Phi)(|\varphi\rangle\langle\varphi|)|i\alpha\rangle\nonumber\\
& =& \langle\alpha| \tilde{\Theta}(\Phi)(|i\rangle\langle i|)|\alpha\rangle\nonumber\\
& =& \langle\alpha| \Delta^O \circ \Phi \circ \Delta^I (|i\rangle\langle i|)|\alpha\rangle\nonumber\\
& =& \langle\alpha| \Delta^O \circ \Phi  (|i\rangle\langle i|)|\alpha\rangle\nonumber\\
& =& \langle\alpha| \Delta ( \sum_{\beta,\gamma} \langle \beta|\Phi  (|i\rangle\langle i|)|\gamma\rangle |\beta\rangle\langle \gamma|)|\alpha\rangle\nonumber\\
& =& \langle\alpha| (\sum_{\beta} \langle \beta|\Phi  (|i\rangle\langle i|)|\beta\rangle |\beta\rangle\langle \beta|)|\alpha\rangle\nonumber\\
& =&  \langle \alpha|\Phi  (|i\rangle\langle i|)|\alpha\rangle\nonumber\\
& =&  \Phi_{ii\alpha\alpha}. \nonumber\\
\end{eqnarray}

Thus,
\begin{equation}\begin{aligned}
C_{\tilde{\Theta}(\Phi)}
&=[\mathbb{I}\otimes(\Delta^{O}\circ\Phi\circ\Delta^{I})]|\varphi\rangle\langle\varphi| \\
&=\sum_{i, \alpha=0}^{d-1} \frac{\Phi_{i  i\alpha \alpha}}{d}|i \alpha\rangle\langle i \alpha|.
\end{aligned}\end{equation}

Obviously the Kraus operators of $\Theta$ are  $\{M_{i\alpha}=|i\alpha\rangle\langle i\alpha|\}^{d-1}_{i,\alpha=0}$, because

\begin{equation}\begin{aligned}
\Theta\left(C_{\Phi}\right)&=\sum_{i\alpha}M_{i\alpha}C_{\Phi}M_{i\alpha}^{\dagger} \\
&=\sum_{i\alpha}|i\alpha\rangle\langle i\alpha|\left(\sum_{i', j', \alpha', \beta'=0}^{d-1} \frac{\Phi_{i' j'\alpha' \beta'}}{d} |i'\alpha'\rangle\langle j'\beta'|\right)|i\alpha\rangle\langle i\alpha|\\
&=\sum_{i,\alpha=0}^{d-1} \frac{\Phi_{i i \alpha \alpha}}{d}|i\alpha\rangle\langle i\alpha|\\
&=\sum_{i, \alpha=0}^{d-1}\langle i \alpha|C_{\Phi}| i \alpha\rangle|i \alpha\rangle\langle i \alpha|\\
&=C_{\tilde{\Theta}(\Phi)}.
\end{aligned}\end{equation}

It is easy to check that $\Theta$ is a resource destroying map \cite{yingshe}. The essence of the quantum operation   $\Theta$  acting on any Choi-Jamio{\l}kowski state is that all the non-diagonal elements in the selected base are completely eliminated and a new Choi-Jamio{\l}kowski state with only diagonal elements is created.

So we get the specific form of the quantum operation $\Theta$ corresponding to the phase-out superoperation $\tilde{\Theta}$ is
\begin{equation}
\Theta(C_{\Phi})=\sum_{i, \alpha=0}^{d-1}\langle i \alpha|C_{\Phi}| i \alpha\rangle|i \alpha\rangle\langle i \alpha|=C_{\tilde{\Theta}(\Phi)}.
\end{equation}
This completes the proof of Theorem 1.1.

{\bf{Theorem 1.2}}. If  a quantum operation $\Phi\in \mathcal{O}(\mathcal{H}^{I}\rightarrow \mathcal{H}^{O})$ is a completely positive trace preserving operation (CPTP), then $\tilde{\Theta}(\Phi)$ is a  CPTP.

{\bf{Proof}}. Obviously, if a quantum operation $\Phi \in \mathcal{O}(\mathcal{H}^{I}\rightarrow \mathcal{H}^{O})$ is a CPTP, then $\Phi$ satisfies
$$
C_{\Phi}\geq0, ~~~~~\mathrm{tr}_{O}(C_{\Phi})
=\frac{\mathbb{I}}{d},
$$
where $d$ is the dimension of the Hilbert space $\mathcal{H}^{I}$.

It is easy to obtain
\begin{equation}\begin{aligned}
\mathrm{tr}_{O}(C_{\Phi})
&=\mathrm{tr}_{O}\left(\frac{1}{d}\sum_{i,j=0}^{d-1}|i\rangle\langle j|\otimes \sum_{\alpha,\beta=0}^{d-1}\Phi_{ij\alpha\beta}
|\alpha\rangle\langle \beta|\right)\\
&=\left(\frac{1}{d}\sum_{i,j=0}^{d-1}|i\rangle\langle j|\right)\mathrm{tr}\left(\sum_{\alpha,\beta=0}^{d-1}\Phi_{ij\alpha\beta}
|\alpha\rangle\langle \beta|\right)\\
&=\sum_{i,j=0}^{d-1}\frac{\sum_{\alpha=0}^{d-1}\Phi_{ij\alpha\alpha}}{d}|i\rangle\langle j|.
\end{aligned}\end{equation}

As $\Phi \in \mathcal{O}(\mathcal{H}^{I}\rightarrow \mathcal{H}^{O})$ is a CPTP, so one must have
\begin{equation}
\sum_{i,j=0}^{d-1}\frac{\sum_{\alpha=0}^{d-1}\Phi_{ij\alpha\alpha}}{d}|i\rangle\langle j|=\frac{\mathbb{I}}{d}.
\end{equation}

The above equation means that
\begin{equation}
\sum_{\alpha=0}^{^{d-1}}\Phi_{ij\alpha\alpha}=\delta_{ij}.
\end{equation}

 The Choi-Jamio{\l}kowski state corresponding to $\tilde{\Theta}(\Phi)$ is
\begin{equation}
C_{\tilde{\Theta}(\Phi)}=\sum_{i, \alpha=0}^{d-1} \frac{\tilde{\Theta}(\Phi)_{i i \alpha \alpha}}{d} |i \alpha\rangle\langle i \alpha|=\sum_{i, \alpha=0}^{d-1} \frac{\Phi_{i i \alpha \alpha}}{d} |i \alpha\rangle\langle i \alpha|.
\end{equation}

So it is easy to obtain
\begin{equation}\begin{aligned}
\mathrm{tr}_{O}(C_{\tilde{\Theta}(\Phi)})
&=\mathrm{tr}_{O}\left(\frac{1}{d}\sum_{i=0}^{d-1}|i\rangle\langle i|\otimes \sum_{\alpha=0}^{d-1}\tilde{\Theta}(\Phi)_{ii\alpha\alpha}
|\alpha\rangle\langle \alpha|\right)\\
&=\left(\frac{1}{d}\sum_{i=0}^{d-1}|i\rangle\langle i|\right)\mathrm{tr}\left(\sum_{\alpha=0}^{d-1}\tilde{\Theta}(\Phi)_{ii\alpha\alpha}
|\alpha\rangle\langle \alpha|\right)\\
&=\sum_{i=0}^{d-1}\frac{\sum_{\alpha=0}^{d-1}\tilde{\Theta}(\Phi)_{ii\alpha\alpha}}{d}|i\rangle\langle i|\\
&=\sum_{i=0}^{d-1}\frac{\sum_{\alpha=0}^{d-1}\Phi_{ii\alpha\alpha}}{d}|i\rangle\langle i|\\
&=\frac{\mathbb{I}}{d}.
\end{aligned}\end{equation}

Thus we have
\begin{equation}
C_{\tilde{\Theta}(\Phi)}\geq0,~~~~~~~~~~~~~~~~\mathrm{tr}_{O}(C_{\tilde{\Theta}(\Phi)})=\frac{\mathbb{I}}{d},
\end{equation}
which means that $\tilde{\Theta}(\Phi)$ is a CPTP. Theorem 1.2 has been proved.

\subsection{The three sets of quantum superoperations}

We now give the three sets of quantum superoperations based on the phase-out superoperation $\tilde{\Theta}$.

Let $\tilde{\Omega}$ be a quantum superoperation. Suppose that $\tilde{\Omega}$ satisfies
\begin{equation}\label{20}\begin{aligned}
\tilde{\Omega}\circ \tilde{\Theta}=\tilde{\Theta}\circ \tilde{\Omega}\circ \tilde{\Theta},
\end{aligned}\end{equation}
where $\circ$ is the composition of superoperations.  The compound superoperation $\tilde{\Omega}\circ \tilde{\Theta}$ and $\tilde{\Theta}\circ \tilde{\Omega}\circ \tilde{\Theta}$ acting on  the any quantum operation $\Phi$, generate  the following quantum operations
\begin{equation}\begin{aligned}
&\tilde{\Omega}\circ\tilde{\Theta}(\Phi)
=\tilde{\Omega}(\tilde{\Theta}(\Phi)),\\
&\tilde{\Theta}\circ\tilde{\Omega}\circ \tilde{\Theta}(\Phi)=\tilde{\Theta}
(\tilde{\Omega}(\tilde{\Theta}(\Phi))).\\
\end{aligned}\end{equation}

Since Eq.(20) holds, so any output operation of the superoperation $\tilde{\Omega}\circ \tilde{\Theta}$ can be regarded as an output operation of the phase-out superoperation $\tilde{\Theta}$. Thus,  all output operations of superoperation $\tilde{\Omega}\circ \tilde{\Theta}$ are incoherent operations. In other words, the set of incoherent operations is
closed under $\tilde{\Omega}$ if $\tilde{\Omega}$ satisfies Eq.(20). Therefore we call  condition Eq.(20) the
nongenerating coherent conditions. The superoperations satisfying this condition are called nongenerating coherent superoperations. Usually,  nongenerating coherent superoperations is also called maximally incoherent superoperations \cite{fenlei1,jiyi2}. The set of maximally incoherent superoperations  is denoted as $\mathcal{MISO}$.

Next, we consider the following dual form of the
nongenerating coherent condition:
\begin{equation}\begin{aligned}
\tilde{\Theta}\circ \tilde{\Omega}=\tilde{\Theta}\circ \tilde{\Omega}\circ \tilde{\Theta}.
\end{aligned}\end{equation}

Think of the output operation $\tilde{\Theta}(\Phi)$ of an input operation $\Phi$,  the free part of  $\Phi$. Apparently, $\tilde{\Omega}$ cannot make use of the resource stored in any input operation $\Phi$ to affect the free part $\tilde{\Theta}(\Phi)$ if $\tilde{\Omega}$ satisfies the condition Eq.(22). We call this condition the nonactivating coherent condition. The superoperations satisfying this condition are called resource nonactivating coherent superoperations. The set of nonactivating coherent superoperations is denoted as $\mathcal{MISO}^{\ast}$.

Another interpretation is that the  superoperations in the set  $\mathcal{MISO}^{\ast}$ never break up a family: members of the same family must be mapped to the same target family (not necessarily the original one).

In general, the nongenerating coherent condition and the nonactivating coherent condition are independent of each other and do not influence each other.

After that, we consider the case where the quantum superoperation satisfy not only the nongenerating coherent condition, but also the nonactivating coherent condition:
\begin{equation}\begin{aligned}
\tilde{\Omega}\circ \tilde{\Theta}
=\tilde{\Theta}\circ\tilde{\Omega}=\tilde{\Theta}\circ\tilde{\Omega}\circ\tilde{\Theta}.
\end{aligned}\end{equation}
We call this condition the exchangeable resource condition. The superoperations satisfying this condition are called
exchangeable resource superoperations. We also call
exchangeable resource superoperations de-phase incoherent superoperations \cite{fenlei1}. The set of de-phase incoherent superoperations  is denoted as $\mathcal{DISO}$.

It is easy to obtain the following conclusions for  the sets $\mathcal{MISO}$, $\mathcal{MISO}^{\ast}$ and $\mathcal{DISO}$.

(1) The relationship between $\mathcal{MISO}$, $\mathcal{MISO}^{\ast}$ and $\mathcal{DISO}$  is $\mathcal{DISO}=\mathcal{MISO}\cap \mathcal{MISO^{\ast}}.$

(2) For any quantum  superoperation $\tilde{\Omega}$, the following relationship exists
\begin{equation}\begin{aligned}
\tilde{\Theta}\circ \tilde{\Omega}\in \mathcal{MISO}, ~~~~~~\tilde{\Theta}\circ \tilde{\Omega}\circ\tilde{\Theta}\in \mathcal{MISO}.
\end{aligned}\end{equation}

(3) If a quantum  superoperation $\tilde{\Omega}\in  \mathcal{MISO}^{\ast}$, then there exists the following relationship
\begin{equation}\begin{aligned}
\tilde{\Theta}\circ \tilde{\Omega}\in \mathcal{MISO^{\ast}}, ~~~~~~~~\tilde{\Omega}\circ \tilde{\Theta}\in \mathcal{MISO^{\ast}}.
\end{aligned}\end{equation}

Furthermore, we also have

{\bf{Theorem 2.1}}. $\mathcal{MISO}$, $\mathcal{MISO}^{\ast}$ and $\mathcal{DISO}$ are closed to
compound operation and convex combination of quantum superoperations.

{\bf{Proof}}. First, we  prove that $\mathcal{MISO}$ is closed to the compound operation and convex combination of quantum superoperations.

For any quantum superoperations
$\tilde{\Omega}_{1}, \tilde{\Omega}_{2}\in  \mathcal{MISO}$, there are
\begin{equation}\begin{aligned}
&\tilde{\Omega}_{1}\circ\tilde{\Theta}=\tilde{\Theta}\circ\tilde{\Omega}_{1}\circ \tilde{\Theta},\\
&\tilde{\Omega}_{2}\circ\tilde{\Theta}=\tilde{\Theta}\circ\tilde{\Omega}_{2}\circ \tilde{\Theta}.\\
\end{aligned}\end{equation}
Thus we have
\begin{equation}\begin{aligned}
(\tilde{\Omega}_{1}\circ\tilde{\Omega}_{2})\circ\tilde{\Theta}
&=\tilde{\Omega}_{1}\circ(\tilde{\Theta}\circ\tilde{\Omega}_{2}\circ \tilde{\Theta})\\
&=\tilde{\Theta}\circ\tilde{\Omega}_{1}\circ \tilde{\Theta}\circ\tilde{\Omega}_{2}\circ \tilde{\Theta}\\
&=\tilde{\Theta}\circ(\tilde{\Omega}_{1}\circ\tilde{\Omega}_{2})\circ \tilde{\Theta}.\\
\end{aligned}\end{equation}
The above equation means that  $\tilde{\Omega}_{1}\circ\tilde{\Omega}_{2} \in \mathcal{MISO}$. Therefore $\mathcal{MISO}$ is closed to the compound operation.

If there is a quantum superoperation $\tilde{\Omega}$ which satisfies
$\tilde{\Omega}=p\tilde{\Omega}_{1}
+(1-p)\tilde{\Omega}_{2}$, where
$0\leq p\leq 1$, then we have
\begin{equation}\begin{aligned}
\tilde{\Omega}\circ \tilde{\Theta}
&=(p\tilde{\Omega}_{1}+(1-p)\tilde{\Omega}_{2})\circ \tilde{\Theta}\\
&=p\tilde{\Omega}_{1}\circ \tilde{\Theta}+(1-p)\tilde{\Omega}_{2}\circ \tilde{\Theta}\\
&=p\tilde{\Theta}\circ\tilde{\Omega}_{1}\circ \tilde{\Theta}+(1-p)\tilde{\Theta}\circ\tilde{\Omega}_{2}\circ \tilde{\Theta}\\
&=\tilde{\Theta}\circ(p\tilde{\Omega}_{1}+(1-p)\tilde{\Omega}_{2})\circ\tilde{\Theta}\\
&=\tilde{\Theta}\circ\tilde{\Omega}\circ\tilde{\Theta}.\\
\end{aligned}\end{equation}
 Hence $\tilde{\Omega}\in  \mathcal{MISO}$. That indicates  that $\mathcal{MISO}$ is closed to the convex combination of quantum superoperations.

Similarly,  one can  prove that $\mathcal{MISO}^{\ast}$  and $\mathcal{DISO}$ are closed to the compound operation and convex combination of quantum superoperations. Therefore Theorem 2.1 holds.

\subsection{The fidelity coherence measure of operations}

How to measure the coherence of operations is very important in the  study of quantum coherence resources. The coherence of operations have  been measured by the relative entropy coherence measure \cite{Bera,Xu}, the $l_{1}$ norm coherence measure \cite{Bera,Xu} and the robustness coherence measure \cite{moni5}. Next we will give another measure,  the fidelity coherence measure of  operations and provide some examples.

{\bf{Definition 2}}. For any two quantum operations $\Phi,\Lambda\in \mathcal{O}(\mathcal{H}^{I}\rightarrow \mathcal{H}^{O})$, the fidelity $F(\Phi,\Lambda)$ of the operations $\Phi$ and $\Lambda$ is defined as
\begin{equation}
F(\Phi,\Lambda)=F(C_{\Phi},C_{\Lambda}),
\end{equation}
where $C_{\Phi}$ and $C_{\Lambda}$ are Choi-Jamio{\l}kowski states corresponding to $\Phi$ and $\Lambda$ respectively, $F(C_{\Phi},C_{\Lambda})$ is the fidelity of $C_{\Phi}$ and $C_{\Lambda}$.

By Ref. \cite{Xu}, we have

{\bf{Lemma}  1}. If $M$ is a coherence measure for quantum states in the Baumgratz-Cramer-Plenio
(BCP) framework \cite{Baumgratz}, then
\begin{equation}
M(\Phi)\equiv M(C_{\Phi}), ~~~~~\Phi\in \mathcal{O}(\mathcal{H}^{I}\rightarrow \mathcal{H}^{O})
\end{equation}
is a coherence measure for quantum operations.

According to the fidelity of quantum operations and Lemma $1$, we can define the fidelity coherence measure of quantum operations as follows.

{\bf Definition 3}. For any quantum operation $\Phi\in \mathcal{O}(\mathcal{H}^{I}\rightarrow\mathcal{H}^{O})$, if the Choi-Jamio{\l}kowski state $C_{\Phi}$  is a pure state, the fidelity coherence measure of the operation $\Phi$  is defined as
\begin{equation}
M_{\mathrm{f}}(\Phi)=\min _{C_{\Lambda} \in \mathfrak{{F_{C}}}} \sqrt{1-F(C_\Phi,C_{\Lambda})},
\end{equation}
where $F(\rho,\sigma)=(\operatorname{tr}(\sqrt{\sqrt{\rho} \sigma \sqrt{\rho}}))^{2}$ is the Uhlmann fidelity \cite{baozhendu}, $\mathfrak{{F_{C}}}$ is the set of pure incoherent states  in the Hilbert space $\mathcal{H}^I\otimes \mathcal{H}^O$. Then we can define convex-roof extended fidelity coherence measure of the operations for  the general case as
\begin{equation}
M_{\mathrm{f}}(\Phi)=\min _{\{p_{n},\Phi_{n}\}} \sum_{n} p_{n} M_{\mathrm{f}}(\Phi_{n}),
\end{equation}
where the minimum is taken over all the ensembles $\{p_{n},\Phi_{n}\}$ realizing $\Phi$, i.e., $\Phi=\sum_{n}p_{n}\Phi_{n}$, for any $n$, $C_{\Phi_{n}}$ is the pure Choi-Jamio{\l}kowski state corresponding to the operation $\Phi_{n}$.

For the general  unitary operations on a single qubit we have the following result.

{\bf Theorem 3.1}. The fidelity coherence measure of the general form of unitary operations
\begin{equation}\begin{aligned}
U=\mathrm{e}^{\mathrm{i}\alpha}\begin{bmatrix}\mathrm{ e}^{-\mathrm{i}\frac{\beta }{2}} & 0  \\ 0 & \mathrm{e}^{\mathrm{i}\frac{\beta }{2}}  \end{bmatrix}
\begin{bmatrix} \mathrm{cos}\frac{\gamma}{2} & -\mathrm{sin}\frac{\gamma}{2} \\ \mathrm{sin}\frac{\gamma}{2} & \mathrm{cos}\frac{\gamma}{2} \end{bmatrix}
\begin{bmatrix} \mathrm{e}^{-\mathrm{i}\frac{\delta }{2}} & 0  \\ 0 & \mathrm{e}^{\mathrm{i}\frac{\delta }{2}}  \end{bmatrix}
\end{aligned}\end{equation}
on a single qubit is
\begin{equation}
M_{\mathrm{f}}(U)=\min\left\{\sqrt{1-\frac{|\mathrm{cos}\frac{\gamma}{2}|^{2}}{2}},~\sqrt{1-\frac{|\mathrm{sin}\frac{\gamma}{2}|^{2}}{2}}\right\}.
\end{equation}
Here $\alpha,\beta,\gamma$ and $\delta$ are any real numbers.

{\bf Proof}. It is easy to derive that  the   general form of unitary operation $U$ on a single qubit can be written as
\begin{equation}\begin{aligned}
U=\begin{bmatrix} a &   -b\\ \mathrm{e}^{2\mathrm{i}\alpha}b^{\ast} & \mathrm{e}^{2\mathrm{i}\alpha}a^{\ast}  \end{bmatrix},\\
\end{aligned}\end{equation}
where
\begin{equation}
a= \mathrm{e}^{\mathrm{i} (\alpha-\frac{\beta}{2}-\frac{\delta}{2})}\mathrm{cos}\frac{\gamma}{2},
~~~~~~b= \mathrm{e}^{\mathrm{i} (\alpha-\frac{\beta}{2}+\frac{\delta}{2})}\mathrm{sin}\frac{\gamma}{2}.
\end{equation}

The Choi-Jamio{\l}kowski state corresponding to the unitary matrix $U$ is
\begin{equation}\begin{aligned}
C_{U}
&=(\mathbb{I} \otimes U)\left(\frac{1}{2}\sum_{i,j=0}^{1}|ii\rangle\langle jj|\right)(\mathbb{I} \otimes U)^{\dagger}\\
&=\frac{1}{2}\begin{bmatrix}
 aa^{\ast}        & \mathrm{e}^{-2\mathrm{i}\alpha}ab        &
 -ab^{\ast}       & \mathrm{e}^{-2\mathrm{i}\alpha}aa  \\
 \mathrm{e}^{2\mathrm{i}\alpha}a^{\ast}b^{\ast} & bb^{\ast} & -\mathrm{e}^{2\mathrm{i}\alpha}b^{\ast}b^{\ast} & ab^{\ast} \\
 -a^{\ast}b       & -\mathrm{e}^{-2\mathrm{i}\alpha}bb        & bb^{\ast}         & -\mathrm{e}^{-2\mathrm{i}\alpha}ab \\
\mathrm{e}^{2\mathrm{i}\alpha}a^{\ast}a^{\ast}  & a^{\ast}b & -\mathrm{e}^{2\mathrm{i}\alpha}a^{\ast}b^{\ast} &  aa^{\ast}\\
\end{bmatrix}\\
&=\frac{1}{\sqrt{2}}(a|00\rangle+\mathrm{e}^{2\mathrm{i}\alpha}b^{\ast}|01\rangle-b|10\rangle+\mathrm{e}^{2\mathrm{i}\alpha}a^{\ast}|11\rangle)\\
&~~~~\times\frac{1}{\sqrt{2}}(a^{\ast}\langle00|+\mathrm{e}^{-2\mathrm{i}\alpha}b\langle01|-b^{\ast}\langle10|+\mathrm{e}^{-2\mathrm{i}\alpha}a\langle11|).\\
\end{aligned}\end{equation}
Because the Choi-Jamio{\l}kowski state $C_U$ is a pure state, the fidelity coherence measure of $U$ is
\begin{equation}
M_{\mathrm{f}}(U)=\min\left\{\sqrt{1-\frac{|a|^{2}}{2}},~\sqrt{1-\frac{|b|^{2}}{2}}\right\}=\min\left\{\sqrt{1-\frac{|\mathrm{cos}\frac{\gamma}{2}|^{2}}{2}},~\sqrt{1-\frac{|\mathrm{sin}\frac{\gamma}{2}|^{2}}{2}}\right\}.
\end{equation}
The proof of Theorem 3.1 is completed.

Obviously, we have the following result.

{\bf Corollary 3.2}.  For any unitary operation $U$ on a single qubit, the range of values of $M_{\mathrm{f}}(U)$ is $[\frac{\sqrt{2}}{2},\frac{\sqrt{3}}{2}]$.

{\bf Corollary 3.3}.  The  fidelity coherence measure $M_{\mathrm{f}}(\Phi_{\max})$ of the  the  maximally coherent  operation $\Phi_{\max}=\sum_{i,\alpha=0}^{1}\frac{1}{\sqrt{2}}\mathrm{e}^{\mathrm{i}\theta_{i\alpha}}|\alpha\rangle\langle i|$ on a single qubit \cite{Xu},  is largest in the  unitary operations  on a single qubit. Here $\theta_{i\alpha}$ is a real number.

{\bf Proof}. The Choi-Jamio{\l}kowski state corresponding to $\Phi_{\max}$ is
\begin{equation}\begin{aligned}
C_{\Phi_{\max}}=\sum_{i,j,\alpha,\beta=0}^{1}\frac{1}{4}\mathrm{e}^{\mathrm{i} (\theta_{i\alpha}-\theta_{j\beta}})|i\alpha\rangle\langle j\beta|.
\end{aligned}\end{equation}
Because $C_{\Phi_{\max}}$ is a pure state, it is easy to derive that the fidelity coherence measure  of $\Phi_{\max}$ is
\begin{equation}\begin{aligned}
M_{\mathrm{f}}(\Phi_{\max})=\sqrt{1-\frac{1}{4}}=\frac{\sqrt{3}}{2}.
\end{aligned}\end{equation}
By using Corollary 3.2, we know that the  fidelity coherence measure  of the  the  maximally coherent  operation is largest in the   unitary operations  on a single qubit. This ends the proof.

\section{Conclusion}
Based on Choi-Jamio{\l}kowski isomorphism we  investigated  the coherence of quantum operations in the resource theory. We first defined   the  phase-out superoperation and gave  the form of operation which transforms the Choi-Jamio{\l}kowski state of the quantum operation $\Phi$ to the  Choi-Jamio{\l}kowski state of the quantum operation $\tilde{\Theta}(\Phi)$ obtained by  using  the phase-out superoperation $\tilde{\Theta}$ to act on $\Phi$. By using  phase-out superoperation  we defined the set of maximally incoherent superoperations, the set of nonactivating coherent superoperations and the set of de-phase incoherent superoperations and proved that these sets are closed to  compound operation and convex combination of
quantum superoperations.  Further, the fidelity coherence measure of quantum operations was introduced.  The fidelity coherence measure of the unitary operations on the single qubit has been calculated.   We hope  the research  will lead to a better understanding of the coherence of quantum operations.


\begin{thebibliography}{99}




\bibitem{relixue1} M. Lostaglio, D. Jennings, and T. Rudolph, Thermodynamic resource theories, non-commutativity and maximum entropy principles, \href{https://iopscience.iop.org/article/10.1088/1367-2630/aa617f}
    {New J. Phys. \textbf{19}, 043008 (2017)}.


\bibitem{relixue2} F. Brand$\tilde{a}$o, M. Horodecki, N. Ng, J. Oppenheim, and S. Wehner, The second laws of quantum thermodynamics,
    \href{https://www.pnas.org/content/112/11/3275}
    {Proc. Natl. Acad. Sci. USA  \textbf{112}, 3275 (2015)}.

\bibitem{relixue3} P. $\acute{\mathrm{C}}$wikli$\acute{\mathrm{n}}$ski, M. Studzi$\acute{\mathrm{n}}$ski, M. Horodecki, and J. Oppenheim, Limitations on the evolution of quantum coherences: Towards fully quantum second laws of thermodynamics,
    \href{https://journals.aps.org/prl/abstract/10.1103/PhysRevLett.115.210403}
    {Phys. Rev. Lett. \textbf{115}, 210403 (2015)}.


\bibitem{relixue4} A. Misra, U. Singh, S. Bhattacharya, and A. K. Pati, Energy cost of creating quantum coherence,
    \href{https://journals.aps.org/pra/abstract/10.1103/PhysRevA.93.052335}
    {Phys. Rev. A \textbf{93}, 052335 (2016)}.









\bibitem{jisuanji1} H. L. Shi, S. Y. Liu, X. H. Wang, W. L. Yang, Z. Y. Yang, and H. Fan, Coherence depletion in the Grover quantum search algorithm,
    \href{https://journals.aps.org/pra/abstract/10.1103/PhysRevA.95.032307}
    {Phys. Rev. A \textbf{95}, 032307 (2017)}.





\bibitem{jisuanji2} M. Hillery, Coherence as a resource in decision problems: The Deutsch-Jozsa algorithm and a variation,
    \href{https://journals.aps.org/pra/abstract/10.1103/PhysRevA.93.012111}
    {Phys. Rev. A \textbf{93}, 012111 (2016)}.






\bibitem{jisuanji3} A. E. Rastegin, On the role of dealing with quantum coherence in amplitude amplification,
         \href{https://arxiv.org/abs/1703.10118}
         {arXiv:1703.10118}.







\bibitem{jisuanji4} N. Anand and A. K. Pati, Coherence and entanglement monogamy in the discrete analogue of analog grover search,
    \href{https://arxiv.org/abs/1611.04542}
    {arXiv:1611.04542}.



\bibitem{shengwu} V. Giovannetti, S. Lloyd, and L. Maccone, Advances in quantum metrology, \href{https://www.nature.com/articles/nphoton.2011.35}{Nat. Photonics \textbf{5}, 222-229 (2011)}.




\bibitem{Baumgratz} T. Baumgratz, M. Cramer, and M. B. Plenio, Quantifying coherence, \href{https://journals.aps.org/prl/abstract/10.1103/PhysRevLett.113.140401}
    {Phys. Rev. Lett. \textbf{113}, 140401 (2014)}.


\bibitem{fazhan1} J. $\mathrm{{\AA}}$berg, Quantifying superposition, \href{https://arxiv.org/abs/quant-ph/0612146}
 {arXiv:quant-ph/0612146}.



\bibitem{fazhan2}  A. Winter and D. Yang, Operational resource theory of coherence,
    \href{https://journals.aps.org/prl/abstract/10.1103/PhysRevLett.116.120404#fulltext}
    {Phys. Rev. Lett. \textbf{116}, 120404 (2016)}.



\bibitem{fazhan3} Y. Peng, Y. Jiang, and H. Fan, Maximally coherent states and coherence-preserving operations, \href{https://journals.aps.org/pra/abstract/10.1103/PhysRevA.93.032326}
    {Phys. Rev. A  \textbf{93}, 032326 (2016)}.




\bibitem{fazhan4} M. L. Hu, X. Hu, J. Wang, Y. Peng, Y. R. Zhang, and H. Fan, Quantum coherence and geometric quantum discord,  \href{https://www.sciencedirect.com/science/article/pii/S0370157318301893?via%3Dihub}
    {Phys. Rep.  \textbf{1}, 762-764 (2018)}.




\bibitem{fazhan5} A. Streltsov, G. Adesso, and M. B. Plenio, Colloquium: Quantum coherence as a resource, \href{https://journals.aps.org/rmp/abstract/10.1103/RevModPhys.89.041003}
    {Rev. Mod. Phys. \textbf{89}, 041003 (2017)}.

\bibitem{qigaoyan} X. F. Qi, T. Gao, and F. L. Yan, Measuring coherence with entanglement concurrence,
\href{https://doi.org/10.1088/1751-8121/aa7638}{J. Phys. A: Math. Theor. {\bf{50}}, 285301 (2017)}.


\bibitem{fuyangao} L. X. Fu, F. L. Yan, and T. Gao, The block-coherence measures and the coherence measures based on  positive-operator-valued measures,
    \href{https://arxiv.org/abs/2108.04405} {arXiv: 2108.04405}.




\bibitem{zhanggaoyan} L. M. Zhang, T. Gao, and F. L. Yan, Transformations of multilevel coherent states under coherence-preserving operations,
\href{https://doi.org/10.1007/s11433-021-1696-y}
{Sci. China-Phys. Mech.
Astron. {\bf {64}}, 260313 (2021)}.




\bibitem{jiuchan1} C. H. Bennett, A. W. Harrow, D. W. Leung, and J. A. Smolin, On the capacities of bipartite Hamiltonians and unitary gates,
    \href{https://ieeexplore.ieee.org/document/1214070}
    {IEEE Trans. Inf. Theory \textbf{49}, 1895 (2003)}.


\bibitem{jiuchan2} E. Kaur and M. M. Wilde, Amortized entanglement of a quantum channel and approximately teleportation-simulable channels,
    \href{https://www.researchgate.net/publication/318699672_Amortized_entanglement_of_a_quantum_channel_and_approximately_teleportation-simulable_channels}
     {J. Phys. A: Math. Theor. \textbf{51}, 035303
(2017)}.





\bibitem{qufen1} S. Pirandola and C. Lupo, Ultimate precision of adaptive noise estimation,
    \href{}
{Phys. Rev. Lett. \textbf{8}, 100502 (2017)}.

\bibitem{qufen2} M. Berta, C. Hirche, E. Kaur, and M. M. Wilde,
 Amortized channel divergence for asymptotic quantum channel discrimination,
\href{https://arxiv.org/abs/1808.01498}
 {arXiv:1808.01498}.


\bibitem{qufen3} S. Pirandola, R. Laurenza, C. Lupo, and J. L. Pereira, Fundamental limits to quantum channel discrimination,
    \href{https://www.nature.com/articles/s41534-019-0162-y}
    {npj Quantum Inf. \textbf{5}, 50 (2019)}.



\bibitem{moni1} M. Berta, F. G. S. L. Brand$\tilde{a}$o, M. Christandl, and S. Wehner, Entanglement cost of quantum channels,
    \href{https://ieeexplore.ieee.org/document/6556948}
     {IEEE Trans. Inf. Theory \textbf{59}, 6779 (2013)}.




\bibitem{moni2} S. Pirandola, R. Laurenza, C. Ottaviani, and L. Banchi, Fundamental limits of repeaterless quantum communications,
    \href{https://www.nature.com/articles/ncomms15043}
    {Nat. Commun. \textbf{8}, 15043 (2017)}.



\bibitem{moni3} M. M. Wilde, Entanglement cost and quantum channel simulation, \href{https://journals.aps.org/pra/abstract/10.1103/PhysRevA.98.042338}
 {Phys. Rev. A \textbf{98}, 042338 (2018)}.

\bibitem{moni5} M. G. D$\acute{\i}$az, K. Fang, X. Wang, M. Rosati, M. Skotiniotis, J.
Calsamiglia, and A. Winter, Using and reusing coherence to realize quantum processes,
 \href{https://quantum-journal.org/papers/q-2018-10-19-100/}
  {Quantum \textbf{2}, 100 (2018)}.






\bibitem{jiyi1}  D. Rosset, F. Buscemi, and Y. C. Liang, Resource theory of quantum memories and their faithful verification with minimal assumptions,
    \href{https://journals.aps.org/prx/abstract/10.1103/PhysRevX.8.021033}
    {Phys. Rev. X \textbf{8}, 021033 (2018)}.


\bibitem{jiyi2} T. Simnacher, N. Wyderka, C. Spee, X. D. Yu, and O. G\"{u}hne, Certifying quantum memories with coherence,
    \href{https://journals.aps.org/pra/abstract/10.1103/PhysRevA.99.062319}
    {Phys. Rev. A \textbf{99}, 062319 (2019)}.


\bibitem{jiyi3} X. Yuan, Y. Liu, Q. Zhao, B. Regula, J. Thompson, and M. Gu, Universal and operational benchmarking of quantum memories, \href{https://arxiv.org/abs/1907.02521} {arXiv:1907.02521}.




\bibitem{qita1} J. H. Hsieh, S. H. Chen, and C. M. Li, Quantifying quantum-mechanical processes, \href{https://www.nature.com/articles/s41598-017-13604-9}
    {Sci. Rep. \textbf{7}, 13588 (2017)}.



\bibitem{qita2} S. H. Chen, H. Lu, Q. C. Sun, Q. Zhang, Y. A. Chen, and C. M. Li, Discriminating quantum correlations with networking quantum teleportation,
    \href{https://journals.aps.org/prresearch/abstract/10.1103/PhysRevResearch.2.013043}
    {Phys. Rev. Research \textbf{2}, 013043 (2020)}.



\bibitem{qita3} C. C. Kuo, S. H. Chen, W. T. Lee, H. M. Chen, H. Lu, and C. M. Li,
    Quantum process capability,
    \href{https://www.nature.com/articles/s41598-019-56751-x}
    {Sci. Rep. \textbf{9}, 20316 (2019)}.



\bibitem{qita4} E. Wolfe, D. Schmid, A. B. Sainz, R. Kunjwal, and R. W. Spekkens, Quantifying bell: The resource theory of nonclassicality of common-cause boxes,
    \href{https://arxiv.org/abs/1903.06311}
    {arXiv:1903.06311}.






\bibitem{38} Z. W. Liu and A. Winter, Resource theories of quantum channels and the universal role of resource erasure, \href{https://arxiv.org/abs/1904.04201} {arXiv:1904.04201}.
\bibitem{123} G. Gour and A. Winter, How to quantify a dynamical quantum resource, \href{https://journals.aps.org/prl/abstract/10.1103/PhysRevLett.123.150401} {Phys. Rev. Lett. \textbf{123}, 150401 (2019)}.
\bibitem{52} G. Gour and C. M. Scandolo, The entanglement of a bipartite channel, \href{https://arxiv.org/abs/1907.02552} {arXiv:1907.02552}.


\bibitem{Choi1} G. Gour, Comparison of quantum channels by superchannels, \href{https://www.researchgate.net/publication/326912751_Comparison_of_Quantum_Channels_with_Superchannels}
    {IEEE. Trans. Inf. Theory \textbf{65}, 5880-5904 (2019)}.


\bibitem{Choi2} G. Chiribella, G. M. DAriano, and P. Perinotti, Transforming quantum operations: Quantum supermaps,
    \href{https://iopscience.iop.org/article/10.1209/0295-5075/83/30004}
    {Europhys. Lett. \textbf{83}, 30004 (2008)}.






\bibitem{fangxiangyi1} K. F. Bu, A. Kumar, L. Zhang, and J. Wu, Cohering power of quantum operations, \href{https://www.sciencedirect.com/science/article/pii/S0375960117302621}
    {Phys. Lett. A \textbf{381}, 1670-1676 (2017)}.





\bibitem{fangxiangyi2} X. Y. Hu, Channels that do not generate coherence, \href{https://journals.aps.org/pra/pdf/10.1103/PhysRevA.94.012326}
{Phys. Rev. A \textbf{94}, 012326 (2016)}.



\bibitem{fangxiangyi3} K. B. Dana, M. G. D$\acute{\i}$az, M. Mejatty, and A. Winter, Resource theory of coherence: Beyond states,
    \href{https://journals.aps.org/pra/abstract/10.1103/PhysRevA.95.062327}
    {Phys. Rev. A \textbf{95}, 062327 (2017)}.


\bibitem{zhijie1} C. Datta, S. Sazim, A. K. Pati, and P. Agrawal, Coherence of quantum channels, \href{https://www.sciencedirect.com/science/article/pii/S0003491618302264?via%3Dihub}
    {Ann. Phys. (NY) \textbf{397}, 243 (2018)}.



\bibitem{zhijie2} Y. Liu and X. Yuan, Operational resource theory of quantum channels, \href{https://arxiv.org/pdf/1904.02680.pdf}{arXiv:1904.02680}.





\bibitem{zhijie3}   T. Theurer, D. Egloff, L. Zhang, and M. B. Plenio, Quantifying operations with an application to coherence, \href{https://journals.aps.org/prl/abstract/10.1103/PhysRevLett.122.190405}
    {Phys. Rev. Lett. \textbf{122}, 190405 (2019)}.






\bibitem{Korzekwa}K. Orzekwa, S. Czach$\acute{\mathrm{o}}$rski, Z. Puchala, and K. $\dot{Z}$yczkowski, Coherifying quantum channels,
\href{https://iopscience.iop.org/article/10.1088/1367-2630/aaaff3}
{New J. Phys. \textbf{20}, 043028 (2018)}.







\bibitem{Bera} M. N. Bera, Quantifying superpositions of quantum evolutions, \href{https://journals.aps.org/pra/abstract/10.1103/PhysRevA.100.042307} {Phys. Rev. A \textbf{100}, 042307 (2019)}.





\bibitem{Xu} J. W. Xu, Coherence of quantum channels, \href{https://journals.aps.org/pra/abstract/10.1103/PhysRevA.100.052311} {Phys. Rev. A \textbf{100}, 052311 (2019)}.


\bibitem{fenlei1} G. Saxena, E. Chitambar, and G. Gour, Dynamical resource theory of quantum coherence,
    \href{https://www.researchgate.net/publication/336230068_Dynamical_Resource_Theory_of_Quantum_Coherence}
    {Phys. Rev. Research \textbf{2}, 023298 (2020)}.




\bibitem{Choi3} M. D. Choi, Completely positive linear maps on complex matrices,
\href{https://www.sciencedirect.com/science/article/pii/0024379575900750?via%3Dihub}
 {Linear Algebra Appl. \textbf{10}, 285 (1975)}.


\bibitem{Choi4} A. Jamio{\l}kowski, Linear transformations which preserve trace and positive semidefiniteness of operators,
    \href{https://www.sciencedirect.com/science/article/abs/pii/0034487772900110?via%3Dihub}
    {Rep. Math. Phys. \textbf{3}, 275 (1972)}.




\bibitem{yingshe} Z. W. Liu, X. Hu, and S. Lloyd, Resource destroying maps,
                 \href{https://journals.aps.org/prl/abstract/10.1103/PhysRevLett.118.060502}
                 {Phys. Rev. Lett. \textbf{118}, 060502 (2017)}.


\bibitem{baozhendu} A. Uhlmann, The "transition probability" in the state space of a $\ast$-algebra,
    \href{https://www.sciencedirect.com/science/article/abs/pii/0034487776900604}{Rep. Math. Phys. \textbf{9}, 273-279 (1976)}.



\end{thebibliography}
\end{document}